\documentclass{aa}  

\usepackage{graphicx}
\usepackage{txfonts}
\usepackage{xcolor}
\usepackage{hyperref}
\hypersetup{colorlinks, linkcolor=blue, citecolor=blue, urlcolor=blue}
\usepackage{amsmath}
\usepackage{ulem}

\newcommand{\muvec}{\mbox{\boldmath $\mu$}}

\newcommand{\thetavec}{\mbox{\boldmath $\theta$}}
\newcommand{\te}{t_{\rm E}}
\newcommand{\thetae}{\theta_{\rm E}}

\newcommand{\pie}{\pi_{\rm E}}

\newcommand{\dl}{D_{\rm L}}

\definecolor{brown}{rgb}{0.59, 0.29, 0.0}
\definecolor{darkgreen}{rgb}{0.0, 0.42, 0.24}
\definecolor{darkblue}{rgb}{0.01, 0.31, 0.59}
\definecolor{darkblue}{rgb}{0.0, 0.25, 0.42}
\definecolor{blue}{rgb}{0.0,0.0,1.0}
\definecolor{green}{rgb}{0.0,1.0,0.0}

\begin{document}

\title{KMT-2018-BLG-1025Lb: microlensing super-Earth planet orbiting a low-mass star}

\author{
     Cheongho~Han\inst{1} 
\and Andrzej~Udalski\inst{2} 
\and Chung-Uk~Lee\inst{3} 
\\
(Leading authors)\\
     Michael~D.~Albrow\inst{4}   
\and Sun-Ju~Chung\inst{3,5}      
\and Andrew~Gould\inst{6,7}      
\and Kyu-Ha~Hwang\inst{3} 
\and Youn~Kil~Jung\inst{3} 
\and Doeon~Kim\inst{1}
\and Hyoun-Woo~Kim\inst{3} 
\and Yoon-Hyun~Ryu\inst{3} 
\and In-Gu~Shin\inst{3} 
\and Yossi~Shvartzvald\inst{8}    
\and Jennifer~C.~Yee\inst{9}      
\and Weicheng~Zang\inst{10}       
\and Sang-Mok~Cha\inst{3,11} 
\and Dong-Jin~Kim\inst{3} 
\and Seung-Lee~Kim\inst{3,5} 
\and Dong-Joo~Lee\inst{3} 
\and Yongseok~Lee\inst{3,11} 
\and Byeong-Gon~Park\inst{3,5} 
\and Richard~W.~Pogge\inst{7}
\and Chun-Hwey Kim\inst{12}
\and Woong-Tae Kim\inst{13}
\\
(The KMTNet Collaboration),\\
     Przemek~Mr{\'o}z\inst{2,14} 
\and Micha{\l}~K.~Szyma{\'n}ski\inst{2}
\and Jan~Skowron\inst{2}
\and Rados{\l}aw~Poleski\inst{2} 
\and Igor~Soszy{\'n}ski\inst{2}
\and Pawe{\l}~Pietrukowicz\inst{2}
\and Szymon~Koz{\l}owski\inst{2} 
\and Krzysztof~Ulaczyk\inst{15}
\and Krzysztof~A.~Rybicki\inst{2}
\and Patryk~Iwanek\inst{2}
\and Marcin~Wrona\inst{2}
\\
(The OGLE Collaboration)\\
}

\institute{
     Department of Physics, Chungbuk National University, Cheongju 28644, Republic of Korea  \\ \email{\color{blue} cheongho@astroph.chungbuk.ac.kr}     
\and Astronomical Observatory, University of Warsaw, Al.~Ujazdowskie 4, 00-478 Warszawa, Poland                                                          
\and Korea Astronomy and Space Science Institute, Daejon 34055, Republic of Korea                                                                        
\and University of Canterbury, Department of Physics and Astronomy, Private Bag 4800, Christchurch 8020, New Zealand                                     
\and Korea University of Science and Technology, 217 Gajeong-ro, Yuseong-gu, Daejeon, 34113, Republic of Korea                                           
\and Max Planck Institute for Astronomy, K\"onigstuhl 17, D-69117 Heidelberg, Germany                                                                    
\and Department of Astronomy, The Ohio State University, 140 W. 18th Ave., Columbus, OH 43210, USA                                                       
\and Department of Particle Physics and Astrophysics, Weizmann Institute of Science, Rehovot 76100, Israel                                               
\and Center for Astrophysics $|$ Harvard \& Smithsonian 60 Garden St., Cambridge, MA 02138, USA                                                          
\and Department of Astronomy and Tsinghua Centre for Astrophysics, Tsinghua University, Beijing 100084, China                                            
\and School of Space Research, Kyung Hee University, Yongin, Kyeonggi 17104, Republic of Korea                                                           
\and Department of Astronomy \& Space Science, Chungbuk National University, Cheongju 28644, Republic of Korea                                           
\and Department of Physics \& Astronomy, Seoul National University, Seoul 08826, Republic of Korea                                                       
\and Division of Physics, Mathematics, and Astronomy, California Institute of Technology, Pasadena, CA 91125, USA                                        
\and Department of Physics, University of Warwick, Gibbet Hill Road, Coventry, CV4 7AL, UK                                                               
}
\date{Received ; accepted}

\abstract
{}
{
We aim to find missing microlensing planets 
hidden in the unanalyzed lensing events of previous survey data. 
}
{
For this purpose, we conduct a systematic inspection of high-magnification microlensing events, 
with peak magnifications $A_{\rm peak}\gtrsim 30$, in the data collected from high-cadence 
surveys in and before the 2018 season.  From this investigation, we identify an anomaly in the 
lensing light curve of the event KMT-2018-BLG-1025. The analysis of the light curve indicates 
that the anomaly is caused by a very low mass-ratio companion to the lens.  
}
{
We identify three degenerate solutions, in which 
the ambiguity between a pair of solutions (solutions B) is caused by the previously known 
close--wide degeneracy, and the degeneracy between these and the other solution (solution A) 
is a new type that has not been reported before.  The estimated mass ratio between the planet 
and host is 
$q\sim 0.8\times 10^{-4}$ 
for the solution A and 
$q\sim 1.6\times 10^{-4}$ 
for the solutions B.
From the Bayesian analysis conducted with measured observables, we estimate that the masses 
of the planet and host and the distance to the lens are $(M_{\rm p}, M_{\rm h}, \dl)\sim 
(6.1~M_\oplus, 0.22~M_\odot, 6.7~{\rm kpc})$ for the solution A and $\sim (4.4~M_\oplus, 
0.08~M_\odot, 7.5~{\rm kpc})$ for the solutions B. The planet mass is in the category of a 
super-Earth regardless of the solutions, making the planet the eleventh super-Earth planet,
with masses lying between those of Earth and the Solar system's ice giants, discovered by 
microlensing.
}
{}


\maketitle

\section{Introduction}\label{sec:one}

The microlensing method has unique advantages in detecting some specific populations of planets.
It enables one to detect planets orbiting very faint low-mass stars, which are the most common 
populations of stars in the Galaxy, because of the lensing characteristic that does not depend 
on the luminosity of the planet host.  Another very important advantage of the method is that 
its detection efficiency extends to very low-mass planets because of the slow decrease of the 
efficiency with the decrease of the planet/host mass ratio $q$.  The microlensing efficiency 
decreases as $\sqrt{q}$, while the efficiency of other methods, for example, the radial-velocity 
method, decreases in direct proportion to $q$.  See \citet{Gaudi2012} for a review on various 
advantages of the microlensing method. With the sensitivity to planets that are difficult to be 
detected by other methods, microlensing plays an important role to complement other methods not 
only for the complete demographic census of planets but also for the comprehensive understanding 
of the planet formation process.

However, these advantages of the microlensing method, especially the latter one, i.e., the high 
sensitivity to low-mass planets, were difficult to be fully realized during the early generation 
of microlensing experiments, for example MACHO \citep{Alcock1997} and OGLE \citep{Udalski1994}.  
A planetary microlensing signal, in general, appears as a short-term anomaly to the smooth and 
symmetric lensing light curve generated by the host of the planet \citep{Mao1991, Gould1992}. 
For this reason, a microlensing planet search should be carried out in two steps: first by 
detecting lensing events, and second by inspecting planet-induced anomalies in the light curves 
of detected lensing events.  The probability for a star to be gravitationally lensed is very low, 
on the order of $10^{-6}$ for stars located in the Galactic bulge field, toward which microlensing 
surveys have been and are being carried out \citep{Paczynski1991, Griest1991, Sumi2016, Mroz2019}, 
and thus a lensing survey should cover a large area of sky to increase the number of lensing events 
by maximizing the number of monitored stars.  This requirement had limited the cadence of lensing 
surveys, and subsequently the rate of planet detections, especially that of very low-mass planets. 
\citet{Gould1992} proposed to overcome this problem by conducting intensive follow-up observations 
of survey-detected events, which led to the first detections of low mass planets 
\citep{Beaulieu2006, Gould2006}.  However, this approach is necessarily restricted by telescope 
resources to a small number of events.

The planet detection rate has rapidly increased with the operation of high-cadence lensing 
surveys including MOA~II \citep{Bond2001}, OGLE-IV \citep{Udalski2015}, and KMTNet \citep{Kim2016}.
By employing multiple telescopes equipped with large-format cameras, these surveys achieve 
an observation cadence reaching down to 15 min for dense bulge fields.  This cadence is shorter 
than those of the first-generation MACHO and OGLE surveys, that had been carried out with a 
$\sim 1$~day cadence, by about a factor 100.

\begin{table*}[htb]
\small
\caption{Microlensing super-Earth planets\label{table:one}}
\begin{tabular}{lllll}
\hline\hline
\multicolumn{1}{c}{Event}                       &
\multicolumn{1}{c}{$M_{\rm p}$ ($M_\oplus$)}    &
\multicolumn{1}{c}{$M_{\rm host}$ ($M_\odot$)}  &
\multicolumn{1}{c}{Reference}                    \\
\hline
OGLE-2005-BLG-390Lb    &   $\sim 5.5 $  &   $\sim 0.22     $            &  \citet{Beaulieu2006}                       \\
MOA-2007-BLG-192Lb     &   $\sim 3.3 $  &   $\sim 0.06     $            &  \citet{Bennett2008}                        \\
OGLE-2013-BLG-0341Lb   &   $\sim 2   $  &   $\sim 0.1$--0.15            &  \citet{Gould2014}                          \\
OGLE-2016-BLG-1195Lb   &   $\sim 1.25$  &   $\sim 0.067    $            &  \citet{Bond2017}, \citet{Shvartzvald2017}  \\
OGLE-2016-BLG-1928L    &   $\sim 0.3 $  &         --                    &  \citet{Mroz2020}                           \\
OGLE-2017-BLG-1434Lb   &   $\sim 4.4 $  &   $\sim 0.23     $            &  \citet{Udalski2018}                        \\
KMT-2018-BLG-0029Lb    &   $\sim 7.6 $  &   $\sim 1.1      $            &  \citet{Gould2020}                          \\
OGLE-2018-BLG-0532Lb   &   $\sim 8   $  &   $\sim 0.25     $            &  \citet{Ryu2020}                            \\
OGLE-2018-BLG-0677Lb   &   $\sim 4.0$   &   $\sim 0.12     $            &  \citet{Herrera2020}                        \\
KMT-2019-BLG-0842Lb    &   $\sim 10.2$  &   $\sim 0.76     $            &  \citet{Jung2020}                           \\ 
\hline
\end{tabular}
\tablefoot{ 
The sample is selected with a planet mass limit of $M_{\rm p}\lesssim 10~M_\oplus$.
OGLE-2016-BLG-1928L is a free-floating planet, and thus the host mass is not included.
}
\end{table*}

The great shortening of the observation cadence resulted in a rapid increase of the planet 
detection rate. The population of planets with a remarkable increase of the detection rate is 
super-Earth planets, which are defined as planets having masses higher than the mass of Earth, 
but substantially lower than those of the Solar System's ice giants, Uranus and Neptune 
\citep{Valencia2007}.\footnote{ We note that the term "super-Earth" refers only to the mass 
of the planet, and so does not imply anything about the atmosphere structure, surface conditions, 
or size of the planet.} In Table~\ref{table:one}, we list the microlensing super-Earth planets, 
along with the masses of the planets and their hosts, that have been detected from the last 28 
year operation of lensing surveys since 1992.  Among the total ten super-Earth planets, seven 
were detected during the last four years since the full operation of the KMTNet survey in 2016, 
and for all of these events, the KMTNet data played key roles in detecting and characterizing 
the planets.

\begin{figure}
\includegraphics[width=\columnwidth]{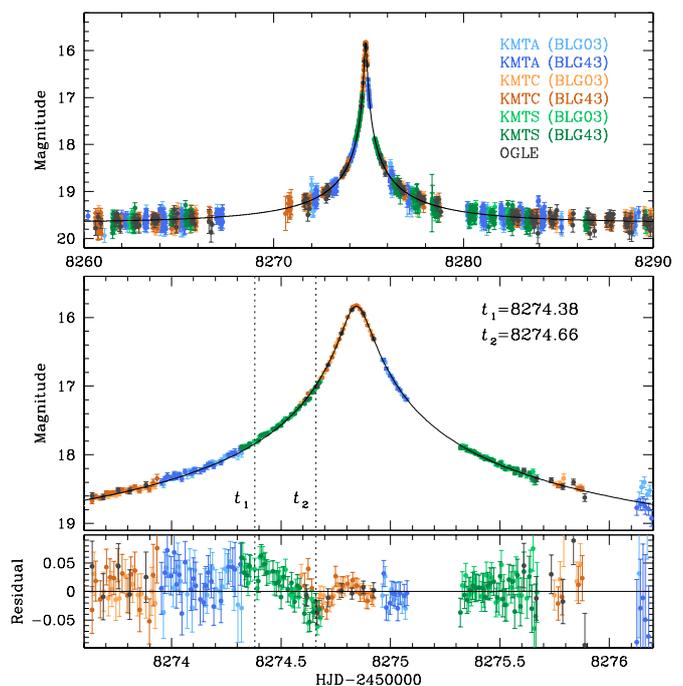}
\caption{
Light curve of KMT-2018-BLG-1025. The upper and lower panels show the whole view and the zoomed-in 
view of the peak region, respectively. The colors of data points indicate the observatories, as 
given in the legend.  The curve plotted over the data points is the 1L1S model, for which the 
residuals in the peak region are presented in the bottom panel.  The dotted vertical lines at 
$t_1\sim 8274.38$ and $t_2\sim 8274.66$ indicate the respective times of the bump and dip in the 
residuals to the 1L1S model.
}
\label{fig:one}
\end{figure}

In this paper, we report the detection of a new microlensing super-Earth planet.  The planet was 
found from a project that has been conducted to search for unrecognized planets in the previous 
KMTNet data collected in and before the 2018 season.  In the first part of this project, 
\citet{Han2020b} investigated lensing events with faint source stars, considering the possibility 
that planetary signals might be missed due to the noise or scatter of data.  From the investigation, 
they found four planetary events that had not been reported before.  The new planetary system that 
we report in this work is found from the second part of the project that has been carried out by 
inspecting subtle planetary signals in the light curves of high-magnification lensing events with 
peak magnifications $A_{\rm peak}\gtrsim 30$.  In this project, high-magnification events are 
selected as targets for reinspection because the sensitivity to planets for these events is high 
\citep{Griest1998}.  Despite the high chance of planet perturbations, some planetary signals produced 
by a non-caustic-crossing channel may not be noticed due to their weak signals \citep{Zhu2014}.

For the presentation of the work, we organize the paper as follows.  The acquisition and 
reduction processes of the data used in the analysis are discussed in Sect.~\ref{sec:two}.  We 
describe the characteristics of the anomaly in the lensing light curve in Sect.~\ref{sec:three}.  
We explain various models tested to explain the observed anomaly, and show that the anomaly is of 
a planetary origin in Sect.~\ref{sec:four}.  The procedure to estimate the angular Einstein radius 
is discussed in Sect.~\ref{sec:five}.  We estimate the physical parameters of the planetary system, 
including the mass and distance, in Sect.~\ref{sec:six}.  We summarize the results and conclude 
in Sect.~\ref{sec:seven}.

\begin{table}[thb]
\small
\caption{Data readjustment factors\label{table:two}}
\begin{tabular*}{\columnwidth}{@{\extracolsep{\fill}}lccc}
\hline\hline
\multicolumn{1}{c}{Data set}                     &
\multicolumn{1}{c}{$N_{\rm data}$}               &
\multicolumn{1}{c}{$k$}                          &
\multicolumn{1}{c}{$\sigma_{\rm min}$}           \\
\hline
KMTA (BLG03)   & 1423  &  1.562     &  0.005     \\
KMTA (BLG43)   & 1246  &  1.483     &  0.010     \\
KMTC (BLG03)   & 1563  &  1.183     &  0.010     \\
KMTC (BLG43)   & 1790  &  1.225     &  0.010     \\
KMTS (BLG03)   & 1357  &  1.208     &  0.005     \\
KMTS (BLG43)   & 1249  &  1.483     &  0.005     \\
OGLE           & 1457  &  1.575     &  0.005     \\
\hline
\end{tabular*}
\tablefoot{$N_{\rm data}$ indicates the number of each data set. }
\end{table}

\section{Observations and data}\label{sec:two}

The planet is found from the analysis of the microlensing event KMT-2018-BLG-1075.  The source star 
of the event lies in the Galactic bulge field with the equatorial coordinates of $({\rm R.A.}, 
{\rm decl.}) =(17:59:27.94, -27:52:41.02)$, which correspond to the Galactic coordinates of $(l, b)
=(2^\circ\hskip-2pt.461, -2^\circ\hskip-2pt.082)$. The flux from the source, which had been constant 
before the lensing-induced magnification with an apparent baseline brightness of $I\sim 19.65$, was 
highly magnified during about 10 days centered at ${\rm HJD}^\prime \equiv {\rm HJD}-2450000\sim 8274.85$.  
The event was found from the post-season inspection of the 2018 season data using the KMTNet Event Finder 
System \citep{Kim2018}.  At the time of finding, the event drew little attention due to the similarity 
of the lensing light curve to that of a regular single-source single-lens (1L1S) event.  See more 
detailed discussion in the following section.

Observations of the event by the KMTNet survey were conducted using three telescopes that are located 
in Australia (KMTA), Chile (KMTC), and South Africa (KMTS). Each telescope has a 1.6m aperture and is 
equipped with a camera yielding 4~deg$^2$ field of view.  The event is located in the two overlapping 
survey fields of ``BLG03'' and ``BLG43'', which are displaced with a slight offset to fill the gaps 
between the chips of the camera. Observations in each field were conducted with a 30~min cadence, 
resulting in a combined cadence of 15~min. Thanks to the high-cadence coverage of the event using 
the globally distributed multiple telescopes, the peak region of the event was continuously and 
densely covered.

\begin{figure}
\includegraphics[width=\columnwidth]{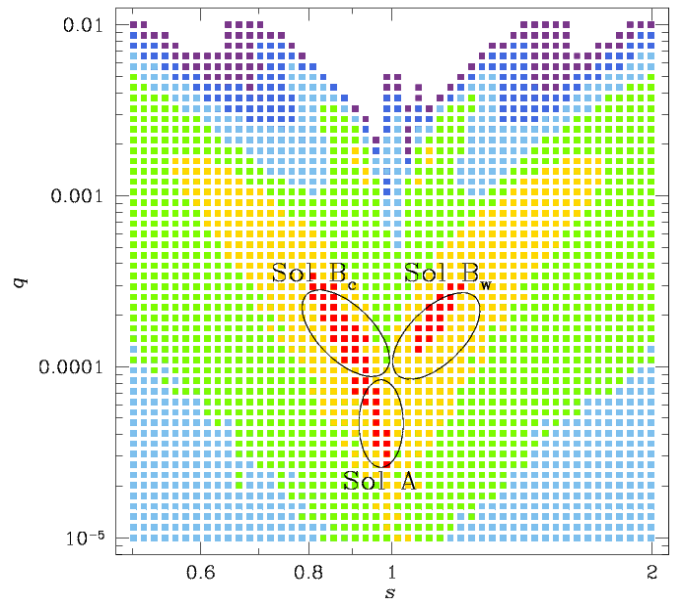}
\caption{
$\Delta\chi^2$ map in the $s$--$q$ plane. The color coding is set to represent points with $<1n\sigma$ 
(red), $<2n\sigma$ (yellow), $<3n\sigma$ (green), $<4n\sigma$ (cyan), $<5n\sigma$ (blue), and $<6n\sigma$ 
(purple), where $n=4$.  The three encircled regions indicate the positions of the three degenerate local 
solutions A, B$_{\rm c}$, and B$_{\rm w}$.
}
\label{fig:two}
\end{figure}

Images of the source were obtained mostly in the $I$ band, and a fraction of images were acquired 
in the $V$ band for the measurement of the source color. Reduction of the data was done using the 
KMTNet pipeline \citep{Albrow2009} based on the difference imaging method \citep{Tomaney1996, 
Alard1998}, that is developed  for the optimal photometry of stars lying in very dense star fields.
For a subset of the KMTC data, an additional photometry was conducted using the pyDIA software 
\citep{Albrow2017} to construct a color-magnitude diagram (CMD) of stars and to measure the color of 
the source star. The detailed procedure of determining the source color is described in Sect.~\ref{sec:five}. 
Error bars of the data estimated from the automatized photometric pipeline were readjusted using the 
method of \citet{Yee2012}.  In this method, the error bars are renormalized by $\sigma=[\sigma_{\rm min}^2 
+ (k\sigma_0)^2]^{1/2}$, where $\sigma_0$ denotes the error estimated from the pipeline, $\sigma_{\rm min}$ 
is a scatter of data, and $k$ is a factor used to make $\chi^2$ per degree of freedom (dof) unity.  In 
Table~\ref{table:two}, we list the numbers and the data readjustment factors for the individual data sets.


Although not alerted at the time of the lensing magnification, the source star of the event lies 
in the field covered by the OGLE survey.  We, therefore, checked the OGLE images containing the 
source and conducted photometry for the source identified by the KMTNet survey.  From this, we 
recover the OGLE photometry data, among which seven data points cover the peak of the light curve.  
OGLE observations were done using the 1.3m telescope of the Las Campanas Observatory in Chile, 
and reduction is carried out using the OGLE photometry pipeline \citep{Udalski2003}.  We publish 
the photometry data to ensure reproducibility of the analysis.  The data are available at 
{\text http://astroph.chungbuk.ac.kr/$\sim$cheongho/data.html}.

\begin{figure*}
\centering
\includegraphics[width=11.5cm]{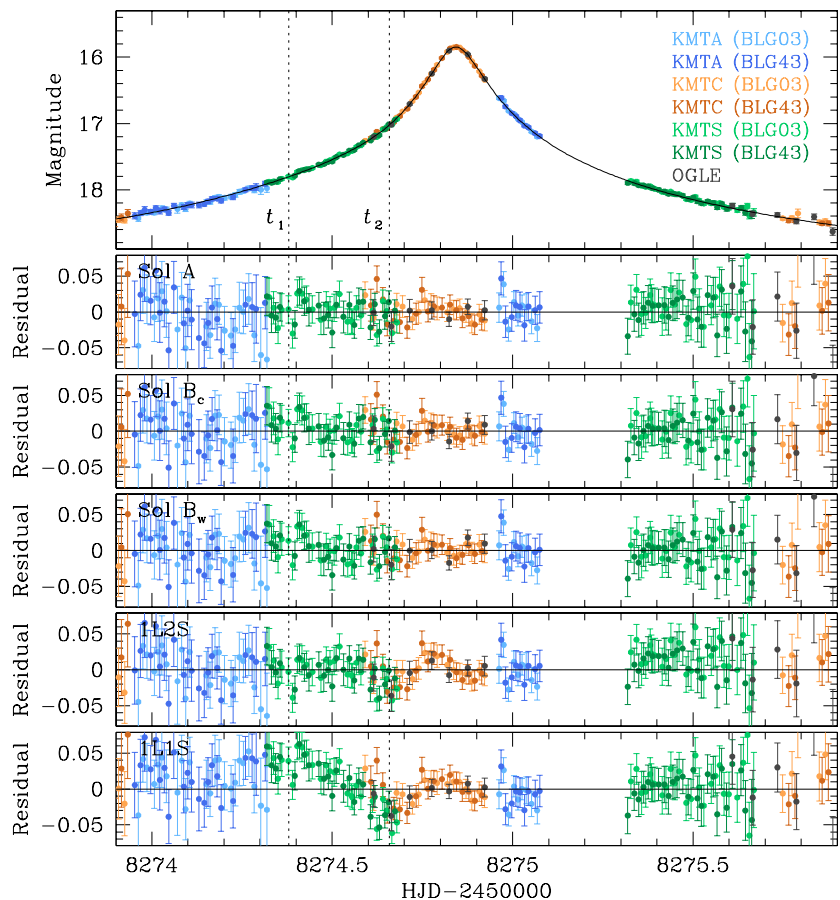}
\caption{
Zoomed-in view of the light curve in the peak region and the residuals from five tested models including 
1L1S, 1L2S, and three 2L1S models (solutions A, B$_{\rm c}$, and B$_{\rm w}$).  Although three 2L1S model 
curves are drawn over the data points in the top panel, it is difficult to distinguish them within the 
line width due to the severity of the degeneracy among the solutions.  
}
\label{fig:three}
\end{figure*}

\section{Characteristics of the anomaly}\label{sec:three}

The light curve of KMT-2018-BLG-1025 is shown in Figure~\ref{fig:one}. At the first glance, it 
appears to have the smooth and symmetric form of a 1L1S event.  A 1L1S modeling yields an impact 
parameter (scaled to the angular Einstein radius $\thetae$) of the lens-source approach and an 
event timescale of $(u_0, \te)\sim (0.0064, 10.1~{\rm days})$, respectively, indicating that the 
event has a relatively short timescale with a high peak magnification of $A_{\rm peak}\sim 1/u_0\sim 150$.  
The 1L1S model curve is plotted over the data points in Figure~\ref{fig:one}.  The full lensing parameters 
and their uncertainties are listed in Table~\ref{table:three}, where $t_0$ indicates the time of the 
closest lens-source approach.  We note that finite-source effects are considered in the 1L1S model, 
but the effects are negligible, and thus the value of the normalized source radius $\rho$ is not 
presented in the table.  The normalized source radius is defined as the ratio of the angular source 
radius $\theta_*$ to $\thetae$, that is, $\rho=\theta_*/\thetae$.

The event was reanalyzed because it was selected in the list of high-magnification events for close 
examinations among the KMTNet events detected in and before the 2018 season in search for planetary 
signals that had not been noticed previously.  From this analysis, we find that the light curve exhibits 
a subtle but noticeable deviation from a 1L1S model.

In the lower two panels of Figure~\ref{fig:one}, we present a zoomed-in view of the light curve 
and residuals from the 1L1S model in the peak region, which shows a slight bump in the residuals 
centered at $t_1\sim 8274.38$ and a dip centered at $t_2\sim 8274.66$.  Although minor, with 
$\Delta I \lesssim 0.05$ magnitude, the deviation drew our attention for two major reasons.  The 
first reason is that the deviation occurred in the central magnification region, in which the chance 
of a planet-induced perturbation is high \citep{Griest1998}.  The second reason is that different 
data sets exhibit a consistent pattern of deviation.  The data sets obtained using the KMTC, located 
in Chile, and KMTS, located in South Africa, telescopes show consistent deviations.  Considering that 
the two telescopes are remotely located, it is difficult to explain the consistency with a coincidental 
systematics in the data such as changes in transparency.  Furthermore, the OGLE data in the deviation 
region exhibit a consistent anomaly pattern with that of the KMTC data, although their coverage is not 
very dense.  Therefore, the deviation is very likely to be real.

\begin{figure}
\includegraphics[width=\columnwidth]{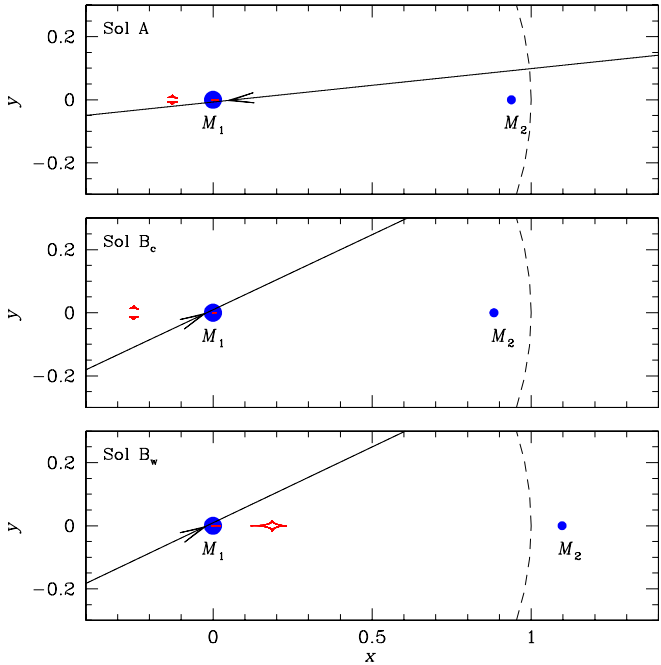}
\caption{
Lens system configurations of the three degenerate 2L1S solutions: A, B$_{\rm c}$, and B$_{\rm w}$.  
In each panel, the blue dots, marked by $M_1$ (host) and $M_2$ (planet), are the lens positions, the 
line with an arrow is the source trajectory, and the cuspy closed curves are the caustics. The dashed 
circle centered at $M_1$ represents the Einstein ring.  The enlarged views in the central magnification 
regions for the individual solutions are presented in Fig.~\ref{fig:five}.  
}
\label{fig:four}
\end{figure}

\section{Interpretation of the anomaly}\label{sec:four}

The fact that the anomaly occurred in the peak region of a high-magnification event suggests the
possibility that the anomaly may be produced by a planetary companion, $M_2$, to the primary lens, 
$M_1$. In order to check this possibility, we conduct an additional modeling under a binary-lens 
(2L1S) interpretation.

The modeling is carried out to find a set of lensing parameters that best explain the observed anomaly 
in the light curve.  In addition to the 1L1S lensing parameters $(t_0, u_0, \te, \rho)$, a 2L1S modeling 
requires one to add three extra lensing parameters of $(s, q, \alpha)$, which represent the projected 
separation (normalized to $\thetae$) and mass ratio between the binary lens components, and the angle 
between the source trajectory and the $M_1$--$M_2$ axis (source trajectory angle), respectively.
The parameter $\rho$ is included to account for potential finite-source effects in the lensing curve 
caused by a source approach close to or a crossing over lensing caustics induced by a lens companion.  
The 2L1S modeling is done in two steps. In the first step, we conduct grid searches for the binary 
lensing parameters $s$ and $q$, while the other parameters are found using a downhill approach based 
on the Markov Chain Monte Carlo (MCMC) algorithm.  Considering that central anomalies can also be 
produced by a very wide or a close binary companion with a mass roughly equal to the primary \citep{Han2009a}, 
we set the ranges of $(s, q)$ wide enough to check the binary origin of the anomaly: $-1.0\leq \log s \leq 1.0$ 
and $-5.0\leq \log q \leq 1.0$.  In the second step, the individual local solutions found from the first step 
are refined by allowing all parameters (including $s$ and $q$) to vary.

\begin{figure}
\includegraphics[width=\columnwidth]{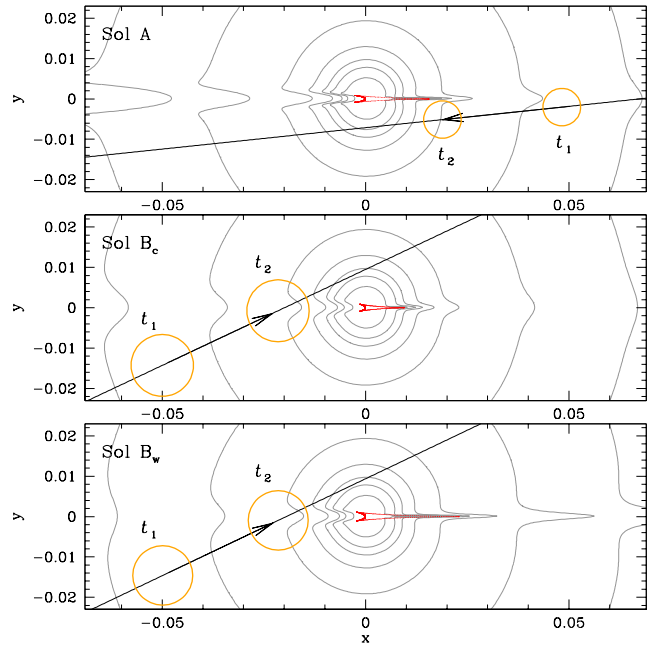}
\caption{
Lens system configurations in the central magnification region for the three 2L1S solutions.  Notations 
are same as those in Fig.~\ref{fig:four}.  The two orange circles represent the source positions at the 
times of the major anomalies at $t_1$ and $t_2$ that are marked in Figs.~\ref{fig:one} and \ref{fig:three}. 
The size of the circle is scaled to the source size.  In the case of the solution A, for which the source 
size cannot be securely measured, the radius of the circle is set to that of the best-fit value. 
}
\label{fig:five}
\end{figure}

\begin{table*}[htb]
\small
\caption{Lensing parameters of various tested models\label{table:three}}
\begin{tabular}{lccccc}
\hline\hline
\multicolumn{1}{c}{Parameter}                      &
\multicolumn{1}{c}{1L1S}                           &
\multicolumn{1}{c}{1L2S}                           &
\multicolumn{3}{c}{2L1S}                           \\
\multicolumn{1}{c}{}                               &
\multicolumn{1}{c}{}                               &
\multicolumn{1}{c}{}                               &
\multicolumn{1}{c}{Solution A}                     &
\multicolumn{1}{c}{Solution B$_{\rm c}$}           &
\multicolumn{1}{c}{Solution B$_{\rm w}$}           \\
\hline
$\chi^2$/dof             &   $10251.0/10082$         &   $10127.9/10077     $   &   $10081.0/10078     $   &    $10089.4/10078     $   &   $10092.8/10078      $  \\
$t_0$ (HJD$^\prime$)     &   $8274.845 \pm 0.001$    &   $8274.845 \pm 0.001$   &   $8274.843 \pm 0.001$   &    $8274.843 \pm 0.001$   &   $8274.843  \pm 0.001$  \\
$u_0$ (10$^{-3}$)        &   $6.416    \pm 0.177$    &   $7.471    \pm 0.532$   &   $7.089    \pm 0.237$   &    $8.575    \pm 0.319$   &   $8.453     \pm 0.334$  \\
$\te$ (days)             &   $10.071   \pm 0.2453$   &   $9.893    \pm 0.262$   &   $9.571    \pm 0.258$   &    $8.889    \pm 0.234$   &   $8.986     \pm 0.239$  \\
$s$                      &   --                      &   --                     &   $0.937    \pm 0.021$   &    $0.883    \pm 0.025$   &   $1.097     \pm 0.036$  \\
$q$ (10$^{-4}$)          &   --                      &   --                     &   $0.829    \pm 0.270$   &    $1.627    \pm 0.499$   &   $1.584     \pm 0.564$  \\
$\alpha$ (rad)           &   --                      &   --                     &   $6.188    \pm 0.009$   &    $2.697    \pm 0.023$   &   $2.693     \pm 0.022$  \\
$\rho$ (10$^{-3}$)       &   --                      &   --                     &   $\lesssim 5.5      $   &    $7.657    \pm 0.645$   &   $7.351     \pm 0.632$  \\ 
$t_{0,2}$ (HJD$^\prime$) &   --                      &   $8274.416 \pm 0.017$   &   --                     &    --                     &   --                     \\
$u_{0,2}$ (10$^{-3}$)    &   --                      &   $7.012    \pm 1.793$   &   --                     &    --                     &   --                     \\
$\rho_2$ (10$^{-3}$)     &   --                      &   --                     &   --                     &    --                     &   --                     \\
$q_F$                    &   --                      &   $0.012    \pm 0.004$   &   --                     &    --                     &   --                     \\
\hline
\end{tabular}
\tablefoot{ 
${\rm HJD}^\prime\equiv {\rm HJD}-2450000$.
}
\end{table*}

From the 2L1S modeling, we identify three degenerate local solutions.  Figure~\ref{fig:two} shows 
the locations of the local solutions in the $\Delta\chi^2$ map on the $s$--$q$ plane obtained from 
the grid search.  The individual locals lie at $(s, q)\sim (0.94, 0.8\times 10^{-4})$, solution ``A'', 
$(0.89, 1.6\times 10^{-4})$, solution ``B$_{\rm c}$'', and $(1.10, 1.6\times 10^{-4})$, solution 
``B$_{\rm w}$''.  Here the subscripts ``c'', standing for close, and ``w'', standing for 
wide, imply that the normalized binary separation is less ($s<1.0$, close solution) and greater 
($s>1.0$, wide solution) than unity, respectively.  The model curves of the individual 2L1S solutions 
and the residuals from the models in the region around the peak of the light curve are shown in 
Figure~\ref{fig:three}.  In Table~\ref{table:three}, we also list the lensing parameters of the 
solutions along with the values of $\chi^2$/dof   for the individual models.  The uncertainty of 
each lensing parameter is estimated as the standard deviation of the distribution of points in the 
MCMC chain under the assumption that the distribution is gaussian.  The $\Delta\chi^2=8.4$ difference 
between the A solution and the minimum of the two B solutions is not big enough to confidently 
distinguish between them.  To be noted among the parameters is that the mass ratios, which are 
$q\sim 0.8\times 10^{-4}$ for the solution~A and $\sim 1.6\times 10^{-4}$ for the solutions~B, are 
very low, indicating that the primary lens is accompanied by a very low-mass planetary companion 
according to the models.  From an additional modeling considering microlens-parallax effects 
\citep{Gould1992b}, we find that it is difficult to securely constrain the microlens parallax 
$\pie$ due to the relatively short timescale, $\sim 9$~days, of the event.

The identified local solutions are subject to two different types of degeneracy. The ambiguity
between the pair of the solutions B$_{\rm c}$ and B$_{\rm w}$ is caused by the well-known close--wide 
degeneracy \citep{Griest1998, Dominik1999, An2005}. The solution A is not subject to this type of 
degeneracy because the source trajectory of the corresponding wide solution passes over the planetary 
caustic located at a position with a separation from $M_1$ of $\sim s-1/s\sim 0.12$ on the planet 
side, and this causes a poor fit of the wide solution to the observed data.

We note that the degeneracy between the A and B solutions is a new type that has not been reported 
before.  The degeneracy is {\it accidental} in the sense that it is caused by the unexpected combination 
of multiple lens parameters instead of being rooted in the lensing physics, for example, the close--wide 
degeneracy that is originated in the invariance of the binary lens equations with $s$ and $s^{-1}$.  
For such accidental degeneracies, it is difficult to identify them from the exploration of the numerous 
combinations of lensing parameters and the simulations of various observational conditions, and thus they 
are mostly identified from the analyses of actual lensing events, as illustrated in the cases of the events 
OGLE-2011-BLG-0526 and  OGLE-2011-BLG-0950/MOA-2011-BLG-336 \citep{Choi2012}, OGLE-2012-BLG-0455/MOA-2012-BLG-206 
\citep{Park2014}, and MOA-2016-BLG-319 \citep{Han2018}.  Although the offsets of 
the source trajectory from the central caustic for both solutions, $\xi \sim u_0/\cos\alpha \sim 7.1 
\times 10^{-3}$ for the solution~A and $\xi \sim 7.7\times 10^{-3}$ for the solution B, are similar 
to each other, this degeneracy is different from the caustic-chirality degeneracy reported by 
\citet{Skowron2018} and \citet{Hwang2018} for two reasons. First, the source stars of the two solutions 
A and B move in almost opposite directions, while the source directions of the two solutions subject to 
the caustic-chirality degeneracy are nearly identical. Second, while the caustic-chirality degeneracy, 
in general, occurs when the source passes a planetary caustic, around which the magnification pattern 
on the left and right sides are approximately symmetric \citep{Gaudi1997}, the magnification pattern 
around the central caustic inducing the observed anomaly is not symmetric \citep{Chung2005}.

The lens system configurations of the individual 2L1S local solutions are shown in Figure~\ref{fig:four}.
In each panel of the figure, the blue dots marked by $M_1$ and $M_2$ denote the positions of the lens 
components, the line with an arrow represents the source trajectory, and the red closed curves are caustics.  
The planet induces two sets of caustics, one lying near the position of $M_1$ (central caustic) and the other 
lying at a position with a separation from $M_1$ of $\sim s-1/s$ (planetary caustic). For all solutions, the 
anomaly is explained by the passage of the source close to the central caustic, but the source incidence 
angles of the solutions A and B differ from one another: $\alpha \sim-5^\circ\hskip-2pt.4$ 
for the solution A and $\alpha \sim 26^\circ$ for the solutions B$_{\rm c}$ and B$_{\rm w}$.

\begin{figure}
\includegraphics[width=\columnwidth]{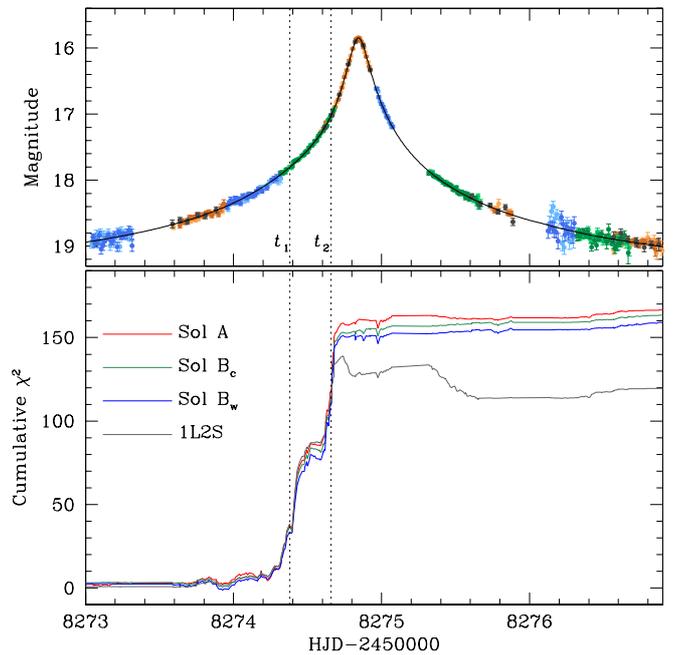}
\caption{
Cumulative distributions of $\Delta\chi^2$ for the three degenerate 2L1S (solutions A, B$_{\rm c}$, and 
B$_{\rm w}$) models and 1L2S model with respect to the 1L1S model. The dotted vertical lines denote the 
times of the major anomalies at $t_1$ and $t_2$ that are marked in Fig.~\ref{fig:one}.  
}
\label{fig:six}
\end{figure}

Figure~\ref{fig:five} shows  the enlarged views of the configuration in the central magnification region 
for the individual solutions.  In each panel, we mark the source positions corresponding to the times 
$t_1$ and $t_2$ (two orange circles), and draw equi-magnification contours (grey curves around the 
caustic).  Around a central caustic, the magnification excess, defined by 
$\epsilon = (A_{\rm 2L1S}-A_{\rm 1L1S})/A_{\rm 1L1S}$, varies depending on the region. Here $A_{\rm 2L1S}$ 
and $A_{\rm 1L1S}$ denote the 2L1S and 1L1S lensing magnifications, respectively.  Positive anomalies occur 
in the regions around the cusps of the caustic, and negative anomalies arise in the outer region of the 
fold caustic and the back end region of the wedge-shaped caustic.  See example maps of magnification 
excess around central caustics presented in \citet{Han2009a} and \citet{Han2009b}. According to the 
solution~A, the bump at $t_1$ is produced when the source passes through the positive excess region 
extending from the protrudent cusp of the central caustic, and the dip at $t_2$ is produced when the source 
moves through the negative excess region formed along the caustic fold. According to the solutions B, on the 
other hand, the bump and dip are produced by the successive passage of the positive and negative excess 
regions formed in the back end region of the caustic, respectively.

Models with the addition of a planetary companion to the lens improves the fit by $\Delta\chi^2\sim 158$ 
-- 170 with respect to the 1L1S solution. To show the region of the fit improvement, we present the 
cumulative distributions of $\Delta\chi^2$ for the three 2L1S solutions in Figure~\ref{fig:six}.  The 
distributions show that the major fit improvement occurs at  around $t_1$ and $t_2$, that are the times 
of the major anomalies from the 1L1S model.  This can also be seen in the residuals of the 2L1S solutions, 
shown in Figure~\ref{fig:three}, which shows that the major residuals from the 1L1S model at around 
$t_1$ and $t_2$ disappear in the residuals of the 2L1S solutions.

\begin{figure}
\includegraphics[width=\columnwidth]{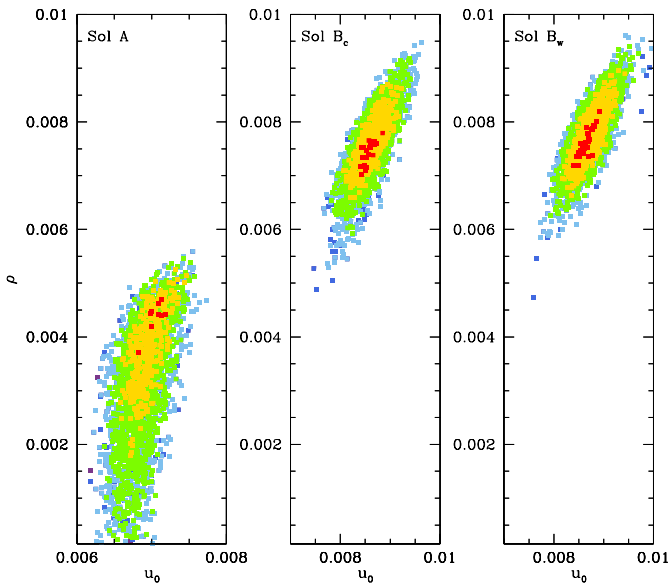}
\caption{
Distributions of points in the MCMC chains for the three degenerate 2L1S solutions. The red, yellow, 
green, cyan, and blue colors represent points within 1$\sigma$, 2$\sigma$, 3$\sigma$, 4$\sigma$, 
and 5$\sigma$, respectively.  The significance level is determined so that $n\sigma$ corresponds to 
$\Delta\chi^2 = n^2$ with rescaled uncertainties.
}
\label{fig:seven}
\end{figure}

We also check the possibility that the anomaly was produced by a companion to the source: 1L2S model.  
Similar to the case of a 2L1S modeling, extra parameters in addition to those of the 1L1S modeling are 
needed for a 1L2S modeling.  Following the parameterization of \citet{Hwang2013}, these extra parameters 
are $(t_{0,2}, u_{0,2}, \rho_2, q_F)$, which denote the time of the closest approach of the second source 
to the lens, the companion-lens separation at that time, the normalized radius of the source companion, 
and the flux ratio between the source stars, respectively.  Considering the possibility that source stars 
approach very close to the lens, we consider finite-source effects in the 1L2S modeling by including two 
parameters ($\rho$, $\rho_2$), which denote the normalized source radii of the first and second source 
stars, respectively.  In the 1L2S modeling, we use the parameters of the 1L1S model as initial parameters, 
and set the other parameters considering the anomaly features of the light curve.  We list the best-fit 
lensing parameters of the 1L2S model in Table~\ref{table:three}, present the residuals from the model in 
Figure~\ref{fig:three}, and show the cumulative $\Delta\chi^2$ distribution with respect to the 1L1S model 
in Figure~\ref{fig:six}.  We note that the normalized source radii of both source stars are not measurable 
due to very week finite-source effects, and thus the values of $\rho$ and $\rho_2$ are not listed in 
Table~\ref{table:three}.  It is found that the 1L2S model reduces the residuals at around $t_1$, but the 
model still leaves noticeable residuals near the peak of the lightcurve.  The fit of the 1L2S model is 
better than the 1L1S model by $\Delta\chi^2\sim 123$, but it is worse than the 2L1S models by 
$\Delta\chi^2\sim 35$ -- 46.  We, therefore, conclude that the anomaly in the lensing light curve was 
generated by a companion to the lens rather than a companion to the source.

\section{Angular Einstein radius}\label{sec:five}

In general cases of lensing events, the angular Einstein radius is estimated from the combination 
of the angular source radius $\theta_*$ and the normalized source radius $\rho$ by
\begin{equation}
\thetae = {\theta_* \over \rho}.
\label{eq1}
\end{equation}
The value of $\theta_*$ can be derived from the color and brightness of the source, and the value 
of $\rho$ is decided from the analysis of the light curve affected by finite-source effects.  Then, 
the prerequisite for the measurement of $\thetae$ is that a lensing light curve should be affected 
by finite-source effects to yield the normalized source radius $\rho$.\footnote{The angular Einstein 
radius can also be measured by separately imaging the lens and source.  By resolving the images, one 
can measure the vector separation $\Delta\thetavec$ between the lens and source and hence their heliocentric 
relative proper motion by $\muvec_{\rm hel}=\Delta\thetavec/\Delta t$, where $\Delta t$ represents the time 
elapsed since the event. Then, the Einstein radius is determined by $\thetae =\mu_{\rm geo}\te$. Here the 
geocentric relative proper motion is related to $\muvec_{\rm hel}$ by $\muvec_{\rm geo}=\muvec_{\rm hel}-
{\bf v}_{\oplus,\perp}\pi_{\rm rel}/{\rm au}$, where ${\bf v}_{\oplus,\perp}$ is Earth's velocity projected 
on the plane of the sky at $t_0$. Due to the long time span $\Delta t$ required for the lens--source 
resolution together with the limited access to high-resolution instrument, there exist just five cases of 
planetary lens events for which the values of $\thetae$ are measured by this method: OGLE-2005-BLG-071 
\citep{Bennett2020}, OGLE-2005-BLG-169 \citep{Batista2015, Bennett2015}, OGLE-2012-BLG-0950 
\citep{Bhattacharya2018}, MOA 2013 BLG-220 \citep{Vandorou2020}, and  MOA-2007-BLG-400 
\citep{Bhattacharya2020}.
}

\begin{figure}
\includegraphics[width=\columnwidth]{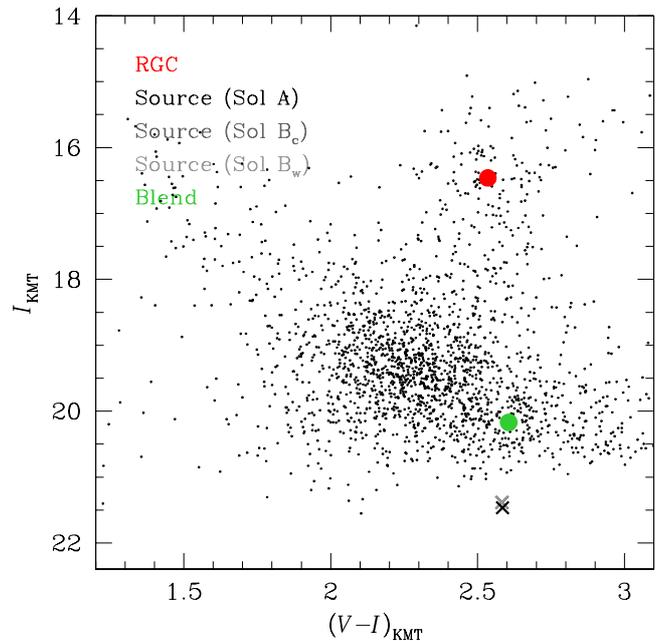}
\caption{
The source (cross mark) position with respect to the centroid of the red giant clump (RGC, red dot) 
in the instrumental color-magnitude diagram of stars lying in the vicinity of the source.  We mark 
source positions corresponding to the three degenerate solutions, which yield very similar source 
locations.  Also marked is the position of the blend (green dot).
}
\label{fig:eight}
\end{figure}

In the case of KMT-2018-BLG-1025, the feasibility of measuring $\rho$ varies depending on the solution. 
We find that finite-source effects are securely detected according to the solutions B$_{\rm c}$ and 
B$_{\rm w}$, but according to the solution A, the effects are not firmly detected and the model is 
consistent with a point-source model within 3$\sigma$.  This is shown in Figure~\ref{fig:seven}, in 
which we present the $\Delta\chi^2$ distributions of points in the MCMC chain obtained from the modeling 
runs of the three degenerate 2L1S solutions.  These scatter plots show that the normalized source radii 
of the B solutions, $\rho\sim 7$ -- $8\times 10^{-3}$, are well determined, but only a upper limit, 
$\rho_{\rm max}\sim 5.5\times 10^{-3}$, can be placed for the solution A.  As a result, the angular Einstein 
radius is determined for the solutions B$_{\rm c}$ and B$_{\rm w}$, but only  a lower limit can be placed 
for the solution~A.  Below, we describe the procedure of $\thetae$ estimation for the individual 
solutions.

\begin{table*}[htb]
\small
\caption{Source color, brightness, Einstein radius, and proper motion\label{table:four}}
\begin{tabular}{llll}
\hline\hline
\multicolumn{1}{c}{Value}                               &
\multicolumn{1}{c}{Solution A}                     &
\multicolumn{1}{c}{Solution B$_{\rm c}$}           &
\multicolumn{1}{c}{Solution B$_{\rm w}$}           \\
\hline
$(V-I, I)$                  &  $(2.585\pm 0.029, 21.465\pm 0.003)$    &   $(2.584\pm 0.027, 21.384\pm 0.002)$   &   $(2.582\pm 0.027, 21.386\pm 0.002)$   \\
$(V-I, I)_{\rm RGC}$        &  $(2.535, 16.461)$                      &   $\leftarrow$                          &   $\leftarrow$                          \\
$(V-I, I)_{{\rm RGC},0}$    &  $(1.060, 14.324)$                      &   $\leftarrow$                          &   $\leftarrow$                          \\
$(V-I, I)_0$                &  $(1.110\pm 0.029, 19.328\pm 0.003)$    &   $(1.109\pm 0.027, 19.247\pm 0.002)$   &   $(1.107\pm 0.027, 19.249\pm 0.002)$   \\
$\theta_*$ (uas)            &  $0.67\pm 0.05$                         &   $0.70\pm 0.05$                        &   $0.69\pm 0.05$                        \\
$\thetae$ (mas)             &  $\geq 0.12$                            &   $0.091\pm 0.007$                      &   $0.094\pm 0.007$                      \\
$\mu$ (mas yr$^{-1}$)       &  $\geq 4.3$                             &   $3.73\pm 0.28$                        &   $3.83\pm 0.29$                        \\
\hline
\end{tabular}
\tablefoot{ 
The notation ``$\leftarrow$'' indicates that the value is same as that presented in the left column.
}
\end{table*}

The angular source radius and the resulting Einstein radius for each solution is estimated following
the routine procedure outlined in \citet{Yoo2004}. In the first step of the procedure, we specify the
source type by placing the positions of the source and the centroid of the red giant clump (RGC)
in the CMD of stars lying in the vicinity of the source.  Figure~\ref{fig:eight} shows the positions 
of the source, marked by a black cross at $(V-I, I)=(2.585\pm 0.029, 21.465\pm 0.003)$, estimated from 
the solution A, and the RGC centroid, red dot at $(V-I, I)_{\rm RGC}=(2.535, 16.461)$, in the instrumental 
CMD constructed using the pyDIA photometry data of the KMTC $I$- and $V$-band images.  We note that the 
source color and brightness estimated from the other solutions, marked by grey crosses and listed in 
Table~\ref{table:four}, result in similar values.  Also marked is the position of the blend, green dot.  
As we will show in the following section, the lens is a very low-mass M dwarf, while the color and 
brightness of the blend correspond to an early main-sequence star or a subgiant.  This implies that 
the contribution of the lens flux to the blended flux is negligible.  We calibrate the source color 
and brightness using the known de-reddened values of the RGC centroid, $(V-I, I)_0=(1.060, 14.324)$ 
\citep{Bensby2013, Nataf2013}, as references. From the measured offsets in the color $\Delta (V-I)$ 
and brightness $\Delta I$ between the source and RGC centroid, the reddening and extinction corrected 
values of the source color and brightness are estimated by
\begin{equation}
(V-I, I)_0 = (V-I, I)_{\rm RGC,0} + \Delta (V-I, I). 
\label{eq2}
\end{equation}

The values of $(V-I, I)_0$ corresponding to the individual solutions are listed in Table~\ref{table:four}. 
The estimated color and brightness are $(V-I, I)_0\sim (1.1, 19.3)$, indicating that the source is an early 
K-type main sequence star. We then convert $V-I$ color into $V-K$ color using the color-color relation of 
\citet{Bessell1988}, and estimate the angular source radius using the $(V-K)$ -- $\theta_*$ relation of 
\citet{Kervella2004}.  The measured source radii are in the range of $0.67\lesssim \theta_*/\mu{\rm as} 
\lesssim 0.70$.  Finally, the angular Einstein radius and the relative lens-source proper motion are estimated 
by the relations $\thetae=\theta_*/\rho$ and $\mu=\thetae/\te$, respectively.

In Table~\ref{table:four}, we list the values of $\theta_*$, $\thetae$, and $\mu$ corresponding to 
the individual solutions.  We note that the lower limits of $\thetae$ and $\mu$ are presented for 
the solution A, for which only the upper limit of $\rho$ is constrained. We note that the Einstein 
radius estimated from the solutions B, $\thetae=0.091$ for the solution B$_{\rm c}$ and 
0.094~mas B$_{\rm w}$, is substantially smaller than $\sim 0.5$~mas of a typical lensing event 
produced by an M dwarf with a mass $\sim 0.3~M_\odot$ located roughly halfway between the lens and 
source.  The angular Einstein radius is related to the lens mass and distance by
\begin{equation}
\thetae = (\kappa M \pi_{\rm rel})^{1/2};\qquad
\pi_{\rm rel} = {\rm au} \left( {1\over D_{\rm L}} - {1\over D_{\rm S}}\right),
\label{eq3}
\end{equation}
where $\kappa=4G/(c^2{\rm au})$ and $D_{\rm S}$ is the distance to the source. Then, the small value of 
$\thetae$ for the solutions B suggests that the lens has a low mass or it is located close to the source.

\section{Physical lens parameters}\label{sec:six}

The lens mass and distance are unambiguously determined by simultaneously measuring $\thetae$ and $\pie$,
which are related to the physical lens parameters by
\begin{equation}
M = {\thetae \over \kappa\pie},\qquad
\dl = {{\rm au}  \over \pie\thetae + \pi_{\rm S} }.
\label{eq4}
\end{equation}
Here $\pi_{\rm S}={\rm au}/D_{\rm S}$ denotes the parallax of the source. In the case of KMT-2018-BLG-1025,
only $\thetae$ is measured for the solutions B$_{\rm c}$ and B$_{\rm w}$, and neither of $\thetae$ and 
$\pie$ is measured for the solution A. Although this makes it difficult to uniquely determine $M$ and $\dl$, 
these parameters can be statistically constrained from a Bayesian analysis with the priors of a lens mass 
function and a Galactic model.

\begin{table}[t]
\small
\caption{Physical lens parameters\label{table:five}}
\begin{tabular*}{\columnwidth}{@{\extracolsep{\fill}}lccc}
\hline\hline
\multicolumn{1}{c}{Parameter}                      &
\multicolumn{1}{c}{Solution A}                     &
\multicolumn{1}{c}{Solution B$_{\rm c}$}           &
\multicolumn{1}{c}{Solution B$_{\rm w}$}           \\
\hline
$M_{\rm h}$ ($M_\odot$)   &  $0.219^{+0.297}_{-0.120}$    &   $0.082^{+0.126}_{-0.044}$    &   $0.084^{+0.131}_{-0.045}$   \\
$M_{\rm p}$ ($M_\oplus$)  &  $6.059^{+8.210}_{-3.321}$    &   $4.444^{+6.808}_{-2.401}$    &   $4.447^{+6.932}_{-2.359}$   \\
$\dl$       (kpc)         &  $6.705^{+0.994}_{-1.166}$    &   $7.470^{+0.908}_{-0.920}$    &   $7.450^{+0.904}_{-0.916}$   \\

$a_\perp$   (au)          &  $1.305^{+0.194}_{-0.227}$    &   $0.508^{+0.062}_{-0.063}$    &   $0.650_{-0.079}^{+0.080}$   \\
\hline
\end{tabular*}
\end{table}

In the Bayesian analysis, we conduct a Monte Carlo simulation to produce artificial lensing events.
For the production of events, we use priors of a mass function, to assign lens masses, and a Galactic
model, to assign lens locations and relative lens-source transverse velocities. For the mass function,
we use a model constructed by combining those of \citet{Zhang2020} and \citet{Gould2000}  mass functions,
and the model considers not only stellar lenses but also substellar brown dwarfs and stellar remnants. 
For the physical lens distribution, we use the modified version of \citet{Han2003} model, in which the 
original double-exponential disk model is replaced with the \citet{Bennett2014} model.  We note that 
the distance to the source, $D_{\rm S}$, is allowed to vary by choosing $D_{\rm S}$ from the physical 
distribution model of the bulge instead of using a fixed value.  For the dynamical 
distribution of the lens and source motion, we adopt the \citet{Han1995} model. A detailed description 
of the adopted priors is given in \citet{Han2020a}.  The number of events produced by the simulation for 
each solution is $10^7$. With the events produced by the simulation, posteriors of $M$ and $\dl$ are 
obtained by constructing the probability distributions of events that are consistent with the measured 
observables.  Although the $\rho$ value is not tightly constrained for the solution~A, we use its 
distribution obtained based on the MCMC links to weight the posteriors of the $M$ and $\dl$.

\begin{figure}
\includegraphics[width=\columnwidth]{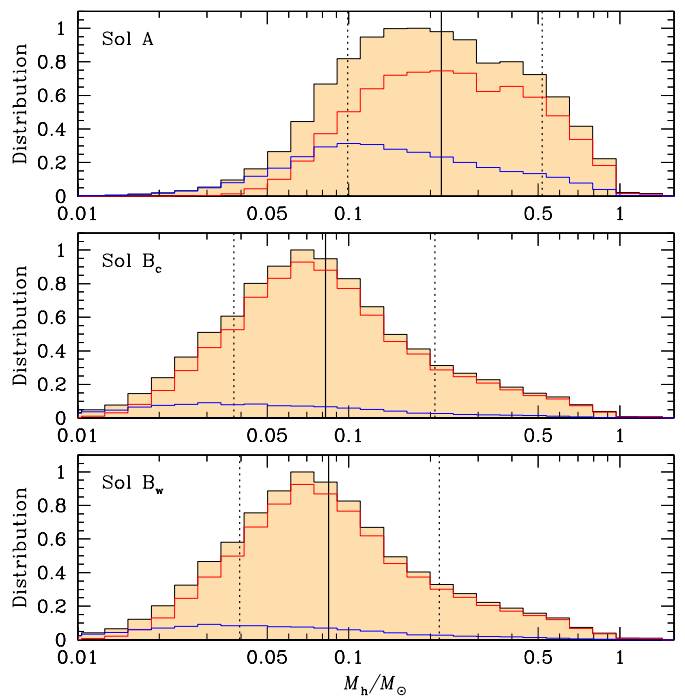}
\caption{
Bayesian posteriors of the host mass for the three degenerate 2L1S solutions. The three curves in each 
panel represent contributions by the disk (blue), bulge (red), and total (black) lens populations. 
The solid vertical line represents the median, and the two dotted vertical lines indicate 1$\sigma$ 
range of the distribution.
}
\label{fig:nine}
\end{figure}

\begin{figure}
\includegraphics[width=\columnwidth]{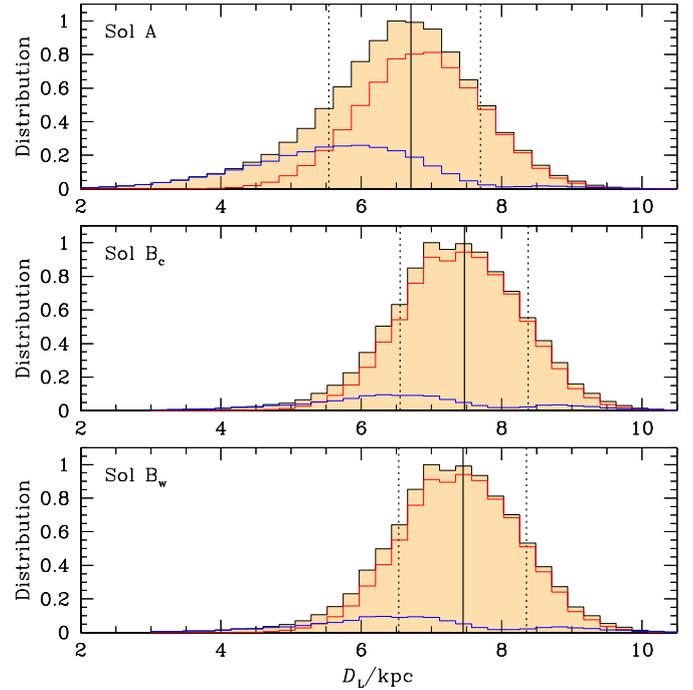}
\caption{
Bayesian posteriors of the lens distance for the three 2L1S solutions. Notations are same as those 
of in Fig.~\ref{fig:nine}.
}
\label{fig:ten}
\end{figure}

The posteriors for the host mass, $M_{\rm h}$, and distance to the lens are shown in Figures~\ref{fig:nine} 
and \ref{fig:ten}, respectively.  For each posterior, we present three distributions, in which the red 
and blue distributions are contributions by the bulge and disk lens populations, respectively, and the black 
distribution is the sum of contributions by the both lens populations. In Table~\ref{table:five}, we list 
the estimated physical parameters of the host and planet ($M_{\rm p}$) masses, distance, and projected 
physical separation ($a_\perp$) of the planet from its host. For each physical parameter, we choose a 
median of the probability distribution as a representative value, and the upper and lower limits are 
estimated as the 16\% and 84\% ranges of the distributions. The estimated masses of the planet and host are 
\begin{equation}
(M_{\rm p}, M_{\rm h}) \sim  
\begin{cases}
(6.1~M_\oplus, 0.22~M_\odot) & \text{for solution A},  \\
(4.4~M_\oplus, 0.08~M_\odot) & \text{for solution B.} \\
\end{cases}
\label{eq5}
\end{equation}
The planet mass is in the category of a super-Earth regardless of the solutions, and thus the planet 
is the eleventh super-Earth planet discovered by microlensing.  The host mass varies depending on the 
solutions: a mid-M dwarf for the solution A and a very late M dwarf or possibly a substellar brown dwarf
for the solutions B.  The estimated distance to the lens is
\begin{equation}
\dl \sim 
\begin{cases}
6.7~{\rm kpc}    & \text{for solution A},  \\
7.5~{\rm kpc}    & \text{for solution B.} \\
\end{cases}
\label{eq6}
\end{equation}
The host mass estimated from the solutions B is substantially smaller than the corresponding value of 
the solution A.  This is because the host mass of the solution A is estimated mostly based on the single 
constraint of the event timescale, $\te\sim 9.6$~days, but the mass of the solutions B is estimated with 
the additional constraint of the small Einstein radius, $\thetae\sim 0.09$~mas.  For the same reason, the 
distance to the lens predicted by the solutions B, $\sim 7.5~{\rm kpc}$, is greater than the distance 
expected from the solution~A, $\sim 6.7~{\rm kpc}$.

The degeneracy between solutions A and B can very likely be lifted if the lens and source can be
resolved from future follow-up observations using high-resolution adaptive optics (AO) instrument.
KMT-2018-BLG-1025 presents an unusual case for which the degeneracy can be lifted purely by
the proper motion measurement, which is the most robust result from AO follow-up observations.
From Figure~\ref{fig:seven}, $\rho_{\rm Sol\ A} <5.5\times 10^{-3}$ and $\rho_{\rm Sol\ B} >5.9
\times 10^{-3}$, both at $3\,\sigma$, while from Tables~\ref{table:three} and \ref{table:four}, 
the quantity $\theta_*/t_{\rm E}$ is about 10\% larger for solution A than solutions B$_{\rm c}$ 
and B$_{\rm w}$. Therefore, the $3\sigma$ limits for $\mu=\theta_*/(\rho t_{\rm E})$ barely overlap. 
Most likely, the actual proper motion measurement will be well away from this boundary, e.g., near 
the best fits $\mu\simeq 3.8~{\rm  mas}~{\rm yr}^{-1}$ (for solutions B) or 
$\mu\simeq 5.7~{\rm  mas}~{\rm yr}^{-1}$ (for solution A). Only if
the measured value is about half way between will the correct solution remain undetermined.
Note that the long tail in the solution A distribution, which prevented a precise estimate of
$\theta_{\rm E}$ and $\mu$ for this case, does not affect the resolution of the degeneracy: if the
true value of $\rho$ is in this tail, then the proper motion will be high, and solution A will be
unambiguously favored. To be confident of detecting the lens, one should allow for proper motions
as low as $\mu\sim 3~{\rm  mas}~{\rm yr}^{-1}$, which are permitted by solutions B. However, even
at this slow pace, the source and lens will be separated by $30$~mas in 2028, the earliest
possible date for first AO light 30m class telescopes. At that point the source and lens can be
easily resolved. By contrast, the close--wide degeneracy between the solutions B$_{\rm c}$ and 
B$_{\rm w}$ cannot be resolved because the relative proper motions expected from the degenerate 
solutions are similar to one another.

\section{Conclusion}\label{sec:seven}

We reported the discovery of a super-Earth planet that was found from the analysis of the lensing
event KMT-2018-BLG-1025. The planetary signal in the lensing light curve had not been noticed 
during the season of the event discovery, and was found from the systematic inspection of high-magnification 
events in the KMTNet data collected in and before the 2018 season.  We identified three degenerate 
solutions, in which the ambiguity between a pair of solutions  was caused by the previously known 
close--wide degeneracy, and the degeneracy between these and the other solution was a new type that 
had not been reported before.  The estimated mass ratio between the planet and host was 
$q\sim 0.8\times 10^{-4}$ for one solution and $\sim 1.6\times 10^{-4}$ for the other pair 
of solutions.  From the Bayesian analysis carried out with the measured observables, we estimated 
that the masses of the planet and host and the distance to the lens were 
$(M_{\rm p}, M_{\rm h}, \dl) \sim (6.1~M_\oplus, 0.22~M_\odot, 6.7~{\rm kpc})$ 
for one solution and $\sim (4.4~M_\oplus, 0.08~M_\odot, 7.5~{\rm kpc})$ for the other solutions.  
The planet mass was in the category of a super-Earth regardless of the solutions, making the planet 
the eleventh super-Earth planet discovered by microlensing.  Due to the substantial difference between 
the relative lens-source proper motions expected from the two sets of solutions, the degeneracy between 
the solutions can be lifted by resolving the lens and source from future high resolution imaging 
observations.  These observations will also yield the mass and distance of the lens, and so the mass 
of the planet.

\begin{acknowledgements}
Work by CH was supported by the grants  of National Research Foundation of Korea 
(2020R1A4A2002885 and 2019R1A2C2085965).
Work by AG was supported by JPL grant 1500811.
This research has made use of the KMTNet system operated by the Korea
Astronomy and Space Science Institute (KASI) and the data were obtained at
three host sites of CTIO in Chile, SAAO in South Africa, and SSO in
Australia.
The OGLE project has received funding from the National Science Centre, Poland, grant
MAESTRO 2014/14/A/ST9/00121 to AU.
\end{acknowledgements}

\end{document}